\documentclass[pre,twocolumn,twoside,byrevtex,superscriptaddress,floatfix]{revtex4-1}

\usepackage{currfile}

\usepackage{graphicx,epsfig,verbatim,enumerate}
\usepackage{amssymb,amsmath}
\usepackage{ifthen}

\usepackage{longtable}

\usepackage{mathtools}

\newboolean{twocolswitch}

\newcommand{\sindex}[1]{}
\newcommand{\nindex}[1]{}

\newcommand{\www}[1]{\url{#1}}

\usepackage{lettrine}

\usepackage{color}

\setboolean{twocolswitch}{true}

\begin{document}

\title{\protect
Quantitative patterns in drone wars
}

\author{
\firstname{Javier}
\surname{Garcia-Bernardo}
}
\email{javier.garcia-bernado@uvm.edu}

\affiliation{Department of Computer Science,
  The University of Vermont,
  Burlington, VT 05401.}

\author{
\firstname{Peter Sheridan}
\surname{Dodds}
}
\email{peter.dodds@uvm.edu}

\affiliation{Department of Mathematics \& Statistics,
  Vermont Complex Systems Center,
  Computational Story Lab,
  \& the Vermont Advanced Computing Core,
  The University of Vermont,
  Burlington, VT 05401.}

\author{
\firstname{Neil F.}
\surname{Johnson}
}
\email{njohnson@physics.miami.edu}

\affiliation{Department of Physics,
University of Miami,
Coral Gables, FL 33124.
}

\date{\today}

\begin{abstract}
  \protect
  Attacks by drones (i.e., unmanned combat air vehicles) continue to
generate heated political and ethical debates.
Here we examine 
the quantitative nature of drone attacks, focusing on how their
intensity and frequency compare with that of other forms of human
conflict. Instead of the power-law distribution found recently for
insurgent and terrorist attacks, the severity of attacks is more akin
to lognormal and exponential distributions, suggesting that the
dynamics underlying drone attacks lie beyond these other forms of
human conflict.
We find that the pattern in the timing of attacks is consistent with one side
having almost complete control, an important if expected result.
We show that these novel features can
be reproduced and understood using a generative mathematical model in
which resource allocation to the dominant side is regulated through a
feedback loop.

\end{abstract}


\maketitle

\section*{Introduction}
Dating back to physicist L.~F.~Richardson's pioneering
work nearly 100 years ago~\cite{Richardson194}, the quantitative
analysis of human conflict has attracted research interest from across
the social, biological, economic, mathematical and physical
sciences~\cite{Morgenstern2013,Cederman2003,Clauset2009,Clauset,Bohorquez2009a,Dedeo2011}.
As in a wide range of other human activities~\cite{Barabasi2005,Gabaix2003}, power
laws have been identified in the severity distribution of individual
attacks in insurgencies and
terrorism~\cite{Johnson2013b,Clauset2009,Clauset,Bohorquez2009a}, and
in the temporal trend in
events~\cite{Johnson2013b,Johnson2011a,Clauset}. These studies found
that across a diverse catalogue of insurgent wars in which a relatively small
opponent such as an insurgency (Red Queen~\cite{Johnson2011a}) fights
a larger one such as a state (Blue King~\cite{Johnson2011a}), the
probability distribution for the severity $s$---the number of fatalities---of an event (i.e., clash
or attack) is given by $P(s) \propto s^{-\alpha}$ where $\alpha \sim
2.5$, while the trend in the timing of attacks is given by $\tau_{n}
= \tau_1 n^{-b}$, where $\tau_n$ is the time interval
between events $n$ and $n+1$, $n=1,2,\dots$ and $b$ is the escalation
parameter. When $b=0$, the Blue King and Red Queen are evenly matched,
with both effectively running on the same spot---hence the terminology
surrounding the Red Queen~\cite{Johnson2011a}. When $b\neq 0$, there
is an escalation in the frequency of attacks which can be interpreted
as a relative advantage between the Red Queen and the Blue King~\cite{Johnson2011a}. The power-law finding for the distribution
of event severities is consistent with the Red Queen (i.e., insurgent
force) evolving dynamically as a self-organizing system composed of
cells (i.e., clusters) that sporadically either fragment under the
pressure of the Blue King (e.g., state) or coalesce to create larger
cells, and taking the severity of attacks as proportional to the sizes
of the resulting cells~\cite{Bohorquez2009a}.

Here, we examine event patterns in the new
form of human conflict offered by unmanned combat air vehicles
(drones)~\cite{wars}. We focus on Pakistan and Yemen because of their
association with drone strike campaigns, using data from the New
America Foundation and the South Asia Terrorism Portal
databases. The situation of drone wars differs from the typical
situation for insurgencies and terrorism in that the attacks are now
carried out by the Blue King on the Red Queen. Moreover, the
sophistication of the action-at-a-distance technology means that any
delay in the Blue King's next attack is likely to have come from a
constraint within Blue itself (e.g., political opposition) as opposed
to any direct counter-adaptation by the Red Queen. Our findings show that drone attacks tend to deviate from the universal patterns observed in the
severity and timing for insurgencies and terrorism, and instead suggest a new regime in which the Blue
King has almost complete control over the conflict. We develop a
generative model in which the timing of attacks is
determined solely by the resources of the Blue King, but are regulated
by a positive feedback loop due to the Blue King's internal
sociopolitical and economic constraints. We show that this simple
model reproduces the main features of the original data 
and hence the unique nature of drone warfare.

\begin{figure*}[tp!]
  \centering	
  \includegraphics[width=\textwidth]{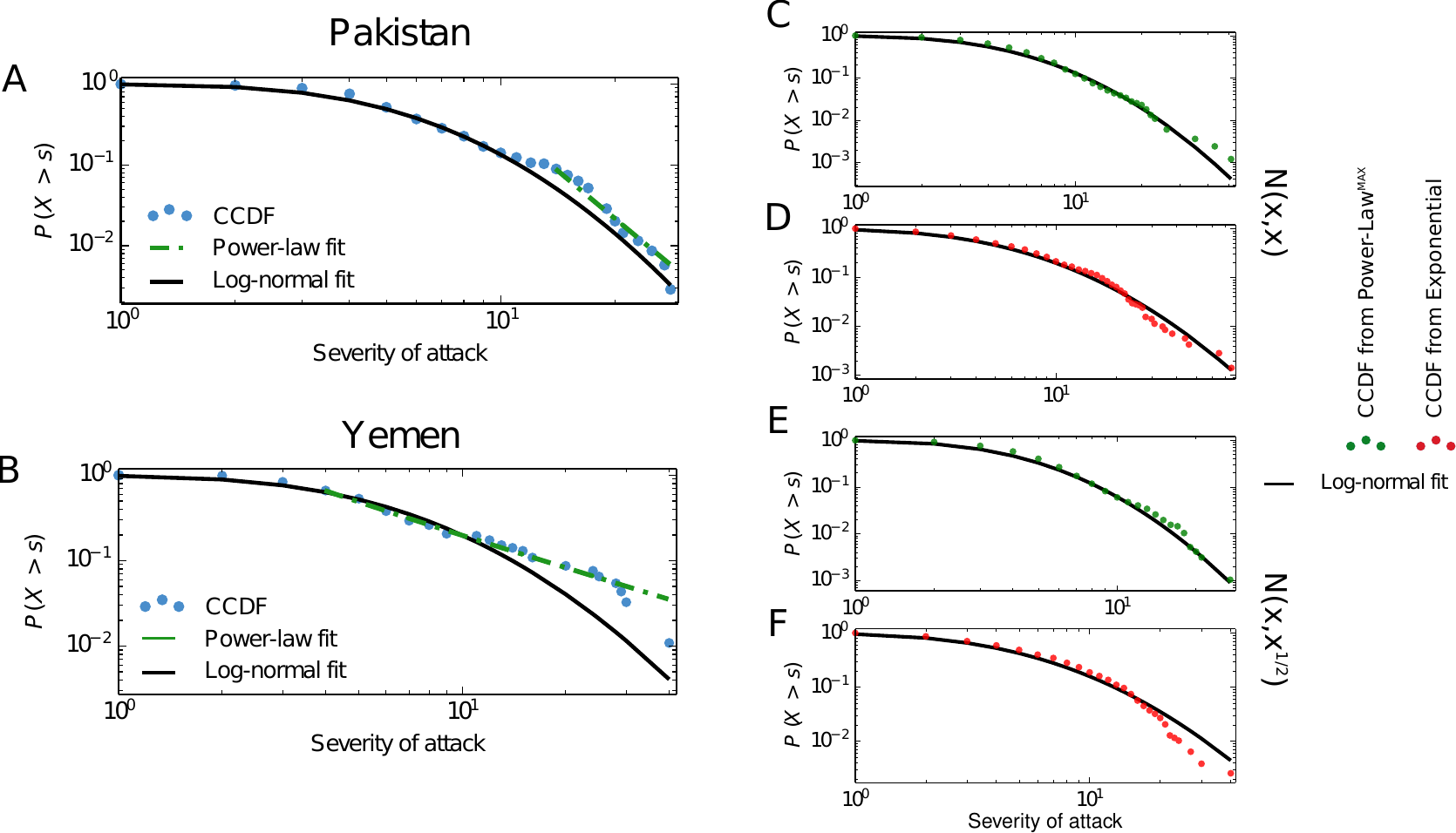}  
  \caption{ 
    \textbf{The severity of drone attacks approximately
      follows a lognormal distribution.}
    Complementary Cumulative Distribution Function (CCDF) of the
    severity of attacks (blue dots and solid line) and best fits to
    power-law (dashed green) and lognormal (solid black) distributions
    for drone attacks in Pakistan (A) and Yemen (B). The optimal
    parameters for each distribution are (A) Power-law: $\alpha =
    4.82$, Log-normal: $\mu = 1.60$ and $\sigma = 0.64$, (B)
    Power-law: $\alpha = 2.21$, Log-normal: $\mu = 1.65$ and $\sigma =
    0.77$.    
    (C-F) CCDFs of the severity of attacks and best fits to log-normal
    distributions. 
    (C and D) The attack size is drawn from a normal distribution
    $N(\mu,\sigma$) with $\mu$ and $\sigma$ corresponding to (C) the
    largest value in 100 random numbers drawn from a power-law
    ($\alpha = 4$) and (D) a random value from a exponential
    distribution ($\lambda = 5$). 
    (E and F) The attack size is drawn from
    a normal distribution with $\mu$ and $\sigma^2$ corresponding to
    (E) the largest value in 100 random numbers drawn from a power-law
    ($\alpha = 4$) and (F) a random value from a exponential
    distribution ($\lambda = 5$). The maximum likelihood parameters
    for the lognormal fits are (C) $\mu = 1.48$ and $\sigma = 0.73$,
    (D) $\mu = 1.51$ and $\sigma = 0.93$, (E) $\mu = 1.33$ and $\sigma
    = 0.63$, (F) $\mu = 1.44$ and $\sigma = 0.86$. }
  \label{fig:drones.data}
\end{figure*}

\begin{figure*}[tp!]
  \centering	
  \includegraphics[width=\textwidth]{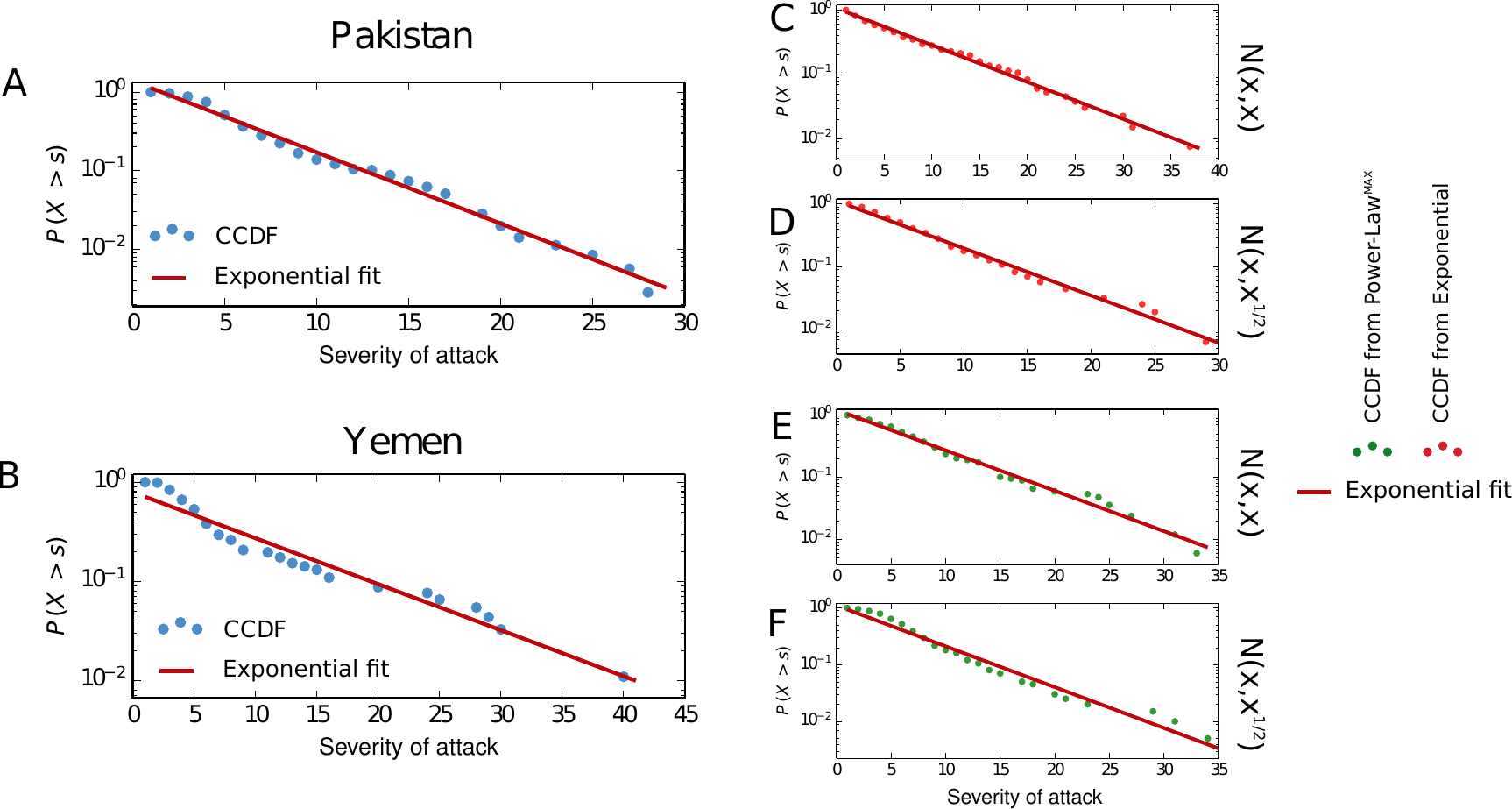}  
  \caption{
    \textbf{The severity of drone attacks approximately follows an exponential distribution.}
    Complementary Cumulative Distribution Function
    (CCDF) of the severity of attacks (blue dots and solid line) and
    best fits to exponential (solid red) distributions for drone attacks in Pakistan (A) and Yemen
    (B). The optimal parameters for each distribution are (A)
    $\lambda = -0.21$ (B) $\lambda = -0.11$.    
    (C-F)  CCDFs of the severity of attacks and best fits to
    exponential distributions. 
    (C and D) The attack size is drawn from a 
    normal distribution with $\mu$ and $\sigma$ ($\sigma^2$ for D)
    corresponding to  a random value from a exponential distribution
    ($\lambda = 5$). 
    (E and F) The attack size is drawn from a normal distribution with $\mu$ and $\sigma$ ($\sigma$ for F) corresponding to the largest value in 100 random numbers drawn from a power-law ($\alpha = 4$). $\lambda$ is measured from the slope of the least squared fit in semi-log scale and corresponds to (C) $\lambda = -0.13$, (D) $\lambda = -0.17$, (E) $\lambda = -0.15$,(F) $\lambda = -0.17$. }
  \label{fig:drones.dataExp}
\end{figure*}

\begin{figure*}[tp!]
  \centering
  \includegraphics[width=\textwidth]{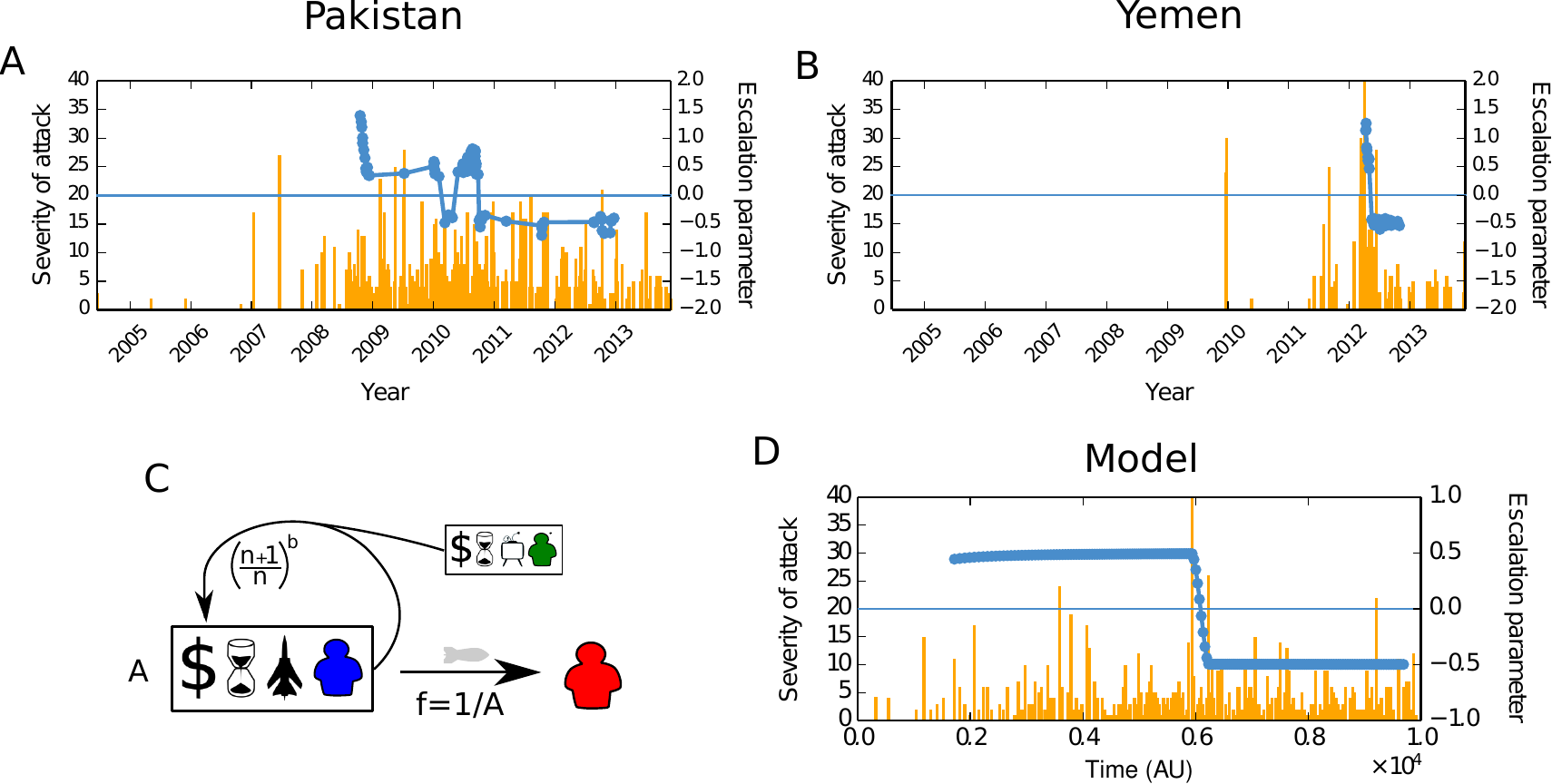}
  \caption{
    \textbf{The timing between attacks reveals power-law relationships.}
    (A and B) The severity of the attacks (vertical lines,
    left axis) and their escalation parameter $b$ (right axis) are plotted for a moving window of 50 attacks in (A) Pakistan and (B) Yemen. (C) A simple model of the process. The Blue King's
    resources (funding, units, experience, etc.) influence the
    frequency of attacks. Resources are invested and create a
    positive feedback loop. The civilian population and other
    variables influence the strength
    of this positive feedback. The amount of resources
    available corresponds to \textit{A}, the advantage of the
    Blue side. 
    (D) Simulated attack severity plot
    using data generated by the model. 
    For the severity of attacks, the size was assumed to be drawn from
    the largest known group, where the group size is distributed as a
    power-law with $\alpha = 4$. 
  }
  \label{fig:drones.model}
\end{figure*}

\section*{Results and Discussion}

Figs.~\ref{fig:drones.data}A--B show the complementary cumulative
distribution function (CCDF) of the severity of drone attacks using
the New America Foundation database. 
We fit power-law and lognormal
distributions (dashed green and solid black lines respectively; see
Methods) for attacks in Pakistan (Figs.~\ref{fig:drones.data}A) and
Yemen (Figs.~\ref{fig:drones.data}B).
We find that the severity of the strikes is approximately described by
lognormal distributions, particularly in the case of Pakistan. In the
case of Yemen, for which we have far less data, the lognormal is more
tentative with the larger events deviating most. This finding of
approximate log-normality is consistent with the notion that a drone
has a specific design and targets (predominantly houses and vehicles)
and hence a pre-determined order of magnitude of the range of
destruction and likely severity. This contrasts with attacks by
terrorist or insurgent clusters whose size and hence lethality crosses
multiple scales, yielding scale-free (power-law) severity
distributions. 

In drone attacks, an approximate lognormal distribution
can arise through at least two mechanisms: First, the fact that the
severity of the attack is the result of many independent processes
(e.g., successful reporting, good visibility, compact target group,
etc.) will itself produce a lognormal distribution in the attack
size. Second, if we take the uncertainty in the casualty number to
scale with the target size, this also produce an approximate lognormal
distribution for many underlying distributions of target sizes. For
example, suppose attacks target the largest known or available Red
group, drawn from a power-law distribution. 
Setting the mean and
standard deviation of a zero-truncated normal distribution to this
value then reproduces an approximate lognormal distribution
(Figs.~\ref{fig:drones.data}C). Similarly, we can imagine that most
attacks target small groups, where the chances of civilian casualties
are lower, and that the probability of targeting larger groups
decreases exponentially. Setting the mean and standard deviation to a
random value from a exponential distribution again yields an
approximate lognormal distribution (Figs.~\ref{fig:drones.data}D). The
same pattern is recovered if the standard deviation is set to scale
with the square root of the mean (Figs.~\ref{fig:drones.data}E--F). We
note that further reduction in the uncertainty of group size increases
the weight of the underlying distribution. For the case where the
largest known group is targeted, this can explain the fat tail
observed for the Yemen data. 

Although we have chosen to focus on fitting lognormal distributions as
the alternative to power laws, other distributions can also provide
good fits. For example, the data agrees well with an exponential
distribution (see Fig.~\ref{fig:drones.dataExp}A--B). Scenarios where
the size of the Red groups is exponentially distributed, as is the
case if the probability of joining a group is constant and independent
of the number of members, would naturally yield exponential
distributions (Fig.~\ref{fig:drones.dataExp}C--D). Approximate
exponential distributions can also be achieved if the groups are
power-law distributed (Fig.~\ref{fig:drones.dataExp}E--F). 
Our purpose
has not been to identify the best alternative to a power-law distribution,
but to show that in contrast with conventional warfare and terrorism,
the data does not follow a power-law distribution and hence feedback
processes are not present across all scales. 

We now turn to the timing of attacks in order to gain insight into the
temporal dynamics of the Blue King-versus-Red Queen
activity. Following previous work~\cite{Johnson2013b}, we plot the
time interval between consecutive attacks $\tau_n$ as a function of
the cardinal number of the attack $n=1,2,3,\dots$. The escalation
parameter $b$ is the exponent of the power-law fit $\tau_n = \tau_1
n^{-b}$, which will be the slope of the best-fit line on a double
logarithmic plot. In the organizational development literature in
which subsequent events are related to production, this is referred to
as a development curve while in the psychology literature, where
subsequent events correspond to completing a certain task, it is
referred to as a learning curve~\cite{Johnson2011a}. In this sense,
the `production' or `completion' of drone attacks has a natural
connection to human activity in these wider fields. For both Pakistan
and Yemen, we find that the parameter $b$ fails to stabilize around
zero (Figs.~\ref{fig:drones.model}A--B), which is the expected value
in a steady state where both sides are adapting well to the opponent's
advances.
Instead, the drone attacks exhibit a large initial escalation (i.e.,
large positive $b$) which then transitions to a de-escalation (i.e.,
large negative $b$). Given the difficulty for a Red Queen without air
defenses to thwart drone attacks directly,
Figs.~\ref{fig:drones.model}A--B suggest that one side (Blue King)
effectively holds complete control for an extended period of time, and
that some internal constraints then arise within the Blue King entity
that eventually de-escalate drone attacks. This is consistent with the
decrease in the escalation rate following the closure of a main drone
base in 2011 ~\cite{benjamin2013drone}. 
Even so, we note that there is some
evidence of Red adaptation to Blue attacks 
as described in a recent report by ``The
Bureau of Investigative Journalism'', 
which shows a decrease in Red
vehicle usage after 2011 corresponding to a peak in attacks on
vehicles \url{http://wherethedronesstrike.com/report/76}. This
suggests that the Red Queen may be able to limit the severity of
attacks.

Fig.~\ref{fig:drones.model}C shows our simple model for explaining
these drone attack patterns. This model is of course over-simplified
given the wealth of unknowns, yet we believe that it is a plausible first
step in explaining the empirical observations. We regard the Blue King
as possessing certain resources, for example experience, units, and
funding. These resources degrade over time if no investment is made in
the Blue King's activity, i.e. if the government does not invest in
its own drone development or information research. We assume that if
there exists investment (i.e., funding, time, etc.) then the available
resources increase due to a positive feedback loop, according to the
escalation observed $A \propto n^b$, where $A$ corresponds to the
advantage of the Blue King over the Red Queen. Similar feedback loops
have been proposed in models of conventional
terrorism~\cite{Clauset2012}, and can be affected by external agents,
for example public opinion or budget changes. For simplicity we take
the frequency of attacks as directly proportional to the resource
level, while the severity of the attack is independent of resources. 

These minimal features are able to replicate the drone strike data
(Figs.~\ref{fig:drones.data}C--F,Figs.~\ref{fig:drones.data}D.
The results are achieved when 
the resources increase as a power-law---hence this is
only sustainable for short periods of time. A constant $b<0$,
corresponding to the frequency of attacks decreasing continuously, is
achieved if the resources decrease continuously, i.e. when there is
little or no investment. Assuming that each drone acts individually,
that the severity of attack varies slowly with the available resources
(which is in turn consistent with some form of adaptation by the Red
side) and that an increase in precision requires significant amounts
of development effort, we are able to recreate approximate lognormal
and exponential distributions for the severity of attacks. 

In summary, our analysis reveals and helps explain patterns in the severity and
timing of attacks in drone wars, which themselves represent a new form
of action-at-a-distance human conflict. We have purposely stepped
aside from issues of ethics or technology, choosing instead to focus
on the event data since they represent a quantitative measure of drone
war impact. We have shown that a simple model, in which the production of
drones evolves from a shared pool of resources controlled by a
feedback loop, is able to recreate the original data and therefore
explain the overall dynamics of the Blue King's drone campaign.
Going forward, our model could be also used to 
explore how wars might unfold when drones are used 
by two or more sides in conflicts.

\section*{Methods}

We obtained all data from the New America database:
\url{http://securitydata.newamerica.net/.newamerica.net/} and crosschecked with the South Asia Terrorism Portal
database:
\url{http://www.satp.org/satporgtp/countries/pakistan/}.

We obtained the best fit to power-law distributions following Clauset et
al.~\cite{Clauset2009}. We fitted lognormal distributions using the
maximum likelihood estimators. For the escalation rate analysis, $\tau_n
= \tau_1 n^{-b}$, we plotted the number of attack vs. the time between
attacks on a log-log scale. We used a rolling window of 50 attacks and accepted every value of $b$ that
allowed for a correlation greater than 20\%, which allows us to
measure fast transitions.

We simulated 200 attacks with our
model. The initial advantage was set to $1$. The time to the
next attack is equal to $323 \cdot A^{-1}$, where the 323 mimics $t_0$ for the Pakistan conflict. At every step (attack) the advantage
of the Blue King changed by the factor 
$\left((n+1)/n\right)^b$, where $n$ is the attack number.
For the first 50 attacks 
$b = 0.5$; 
for the last 50 attacks 
$b = -0.5$.

\acknowledgments
PSD was supported by NSF CAREER Grant No. 0846668.

\clearpage

\end{document}